# A modified two-leaf light use efficiency model for improving the simulation of GPP using a radiation scalar

Xiaobin Guan [a, b], Jing M. Chen [b, c,*], Huanfeng Shen [a, d], Xinyao Xie [e]

[a] *School of Resource and Environmental Sciences, Wuhan University, Wuhan 430079, China.*

[b] *Department of Geography and Planning, University of Toronto, Toronto ON M5S3G3, Canada.*

[c] *School of Geographical Sciences, Fujian Normal University, Fuzhou 350117, China.*

[d] *Collaborative Innovation Center of Geospatial Technology, Wuhan 430079, China.*

[e] *Institute of Mountain Hazards and Environment, Chinese Academy of Sciences, Chengdu 610041, China*

\* Corresponding author: Jing M. Chen (jing.chen@utoronto.ca)

## Abstract

Two-leaf light use efficiency (TL-LUE) models are efficient methods to simulate regional and global gross primary productivity (GPP). A TL-LUE model has previously been shown to outperform the big-leaf MOD17 model through separate consideration of the contributions of sunlit and shaded leaves. However, the impacts of radiation intensity on LUE are inadequately considered in the TL-LUE model and the maximum LUEs of sunlit and shaded leaves are assigned as different constants, which often induce large uncertainties. Therefore, a TL-LUE model modified with a radiation scalar (RTL-LUE) is developed in this paper. The same maximum LUE is used for both sunlit and shaded leaves, and the difference in LUE between sunlit and shaded leaf groups is determined by the same radiation scalar. The RTL-LUE model was calibrated and validated at global 169 FLUXNET eddy covariance (EC) sites. Results indicate that although GPP simulations from the TL-LUE model match well with the EC GPP, the RTL-LUE model can further improve the simulation, for half-hour, 8-day, and yearly time scales. The TL-LUE model tends to overestimate GPP under conditions of high incoming photosynthetically active radiation (PAR), because the radiation-independent LUE values for both sunlit and shaded leaves are only suitable for low-medium (e.g. average) incoming PAR conditions. The errors in the RTL-LUE model show lower sensitivity to PAR, and its GPP simulations can better track the diurnal and seasonal variations of EC GPP by alleviating the overestimation at noon and growing seasons associated with the TL-LUE model. This study demonstrates the necessity of considering a radiation scalar in GPP simulation in LUE models even if the first-order effect of radiation is already considered through differentiating sunlit and shaded leaves. The simple RTL-LUE developed in this study would be a useful alternative to complex process-based models for global carbon cycle research.

## 1. Introduction

Gross primary productivity (GPP), the integral of vegetation photosynthesis by all leaves at the ecosystem scale, is an important component for the terrestrial ecosystem carbon cycle (Beer et al., 2010; Chen et al., 2019b). GPP represents the total amount of atmospheric carbon dioxide fixed by vegetation per unit of space and time, and thus plays a critical role in quantifying the status and changes of the global carbon budget (Falkowski et al., 2000; Marcott et al., 2014). Under the background of the rapid global change, it is of great significance to accurately quantify GPP at regional or global scales, in order to advance our understanding of the interactions between terrestrial ecosystems and changes in atmospheric $CO_2$, temperature, and other variables (Xia et al., 2015; Yang et al., 2013).

In the past decades, numerous models with various complexities and structures have been developed to simulate GPP at the regional or global scale, mainly including process-based models and light use efficiency (LUE) models (Marshall et al., 2018; Yuan et al., 2014). Among them, LUE models are favored in many applications due to their simple model structure and input requirement, making the models easy to use relative to process-based models, which require large numbers of prescribed parameters and input datasets (Dong et al., 2015; Mattos et al., 2020). LUE models can be conveniently driven by the remote sensing vegetation index (VI) data, which are available globally at various resolutions (Guan et al., 2019; Xiao et al., 2010; Zhang et al., 2017). Based on the assumption that GPP linearly increases with the canopy absorbed photosynthetically active radiation (APAR), a variety of big-leaf LUE models were developed, such as the MOD17 model (Running et al., 2004), the eddy covariance (EC-LUE) model (Yuan et al., 2007), the Vegetation Photosynthesis Model (VPM) (Xiao et al., 2004), and the Carnegie‐Ames‐Stanford Approach (CASA) (Potter et al., 1993). The main disparity of these models concentrates on the calculation of APAR and environmental scalars used to constrain the actual LUE departing from the optimum LUE (McCallum et al., 2013; Zhang et al., 2015).

These big-leaf LUE models treat the entire vegetation canopy as a big extended leaf, assuming that the direct and diffuse radiation are absorbed without differences in their use for photosynthesis (Haxeltine and Prentice, 1996; Wang and Leuning, 1998). However, the photosynthesis rates of sunlit and shaded leaves in plant canopies are respectively limited by the maximum carboxylation rate and the electron transport rate, and hence have different LUE values (Chen et al., 1999; De Pury and Farquhar, 1997; Propastin et al., 2012). Different groups of leaves (sunlit and shaded) have different levels of exposure to sunlight, due to their different positions in the canopy (Liu et al., 1997; Oliphant et al., 2011). Sunlit leaves absorb both direct and diffuse radiation, which are easily light-saturated. In contrast, shaded leaves can only absorb diffuse radiation because they are blocked by sunlit leaves, which are usually limited by low APAR (Chen et al., 1999; Mercado et al., 2009;

Rap et al., 2018; Zhang et al., 2011). As a result, there are specific differences in APAR and LUE of the sunlit and shaded leaves, and the same treatment of sunlit and shaded leaves in big-leaf models will induce bias (Propastin et al., 2012; Wang et al., 2015). In fact, it has already been found that the MODIS GPP product underestimates GPP in regions with high vegetation coverage due to insufficient consideration of the contribution of shaded leaves (Chen et al., 2020; He et al., 2018; Zhang et al., 2012). In order to overcome this problem, a two-leaf LUE (TL-LUE) model was developed on the basis of the MOD17 model (He et al., 2013) by integrating the approach in the Boreal Ecosystem Productivity Simulator (BEPS) to separate the LAI and PAR of sunlit and shaded leaves (Chen et al., 1999; Liu et al., 1997). This TL-LUE model has been proved to outperform the big-leaf MOD17 model across a wide range of biomes, which highlights the necessity of separate treatments of sunlit and shaded leaves in LUE models (He et al., 2013; Zhou et al., 2016).

Nevertheless, the difference in LUE between different groups of leaves in the TL-LUE model are totally determined by the independent maximum LUE of sunlit and shaded leaves ($\varepsilon_{msu}$ and $\varepsilon_{msh}$), which are the parameters manually set for different biomes (He et al., 2013; Zhou et al., 2016). The actual LUE of sunlit and shaded leaves are reduced from the optimum ($\varepsilon_{msu}$ and $\varepsilon_{msh}$) by the same scalars of daily minimum temperature and vapor pressure deficit (VPD). However, the differences in LUE between sunlit and shaded leaves should mainly be attributed to the different intensity of radiation they received, and the maximum LUE is almost the same for the two groups of leaves (Chen et al., 1999; De Pury and Farquhar, 1997; Liu et al., 1997). The maximum LUE of a leaf is only decided by the physiological traits of the leaf itself, which should be nearly the same whether or not it is sunlit or shaded (Koyama and Kikuzawa, 2010; Leverenz, 1987). Although there are small differences in nitrogen content and carboxylation capacity between sunlit and shaded leaves because plants tend to allocate more resources to leaves at the top, which receive direct radiation more frequently than those at lower layers (Chen et al., 2012), the impacts of these differences on LUE would be much smaller than that of radiation (Hikosaka, 1996). So the similar physiological traits of sunlit and shaded leaves will certainly result in the similar maximum LUE, and the differences of LUE in them should be mostly caused by the different environmental conditions (Farquhar et al., 1980; Haxeltine and Prentice, 1996). Even though slight differences should exist between sunlit and shaded leaves in their temperature and humidity environments, the large difference in the light intensity between them is the dominant factor on their LUE (Leverenz, 1987). A light response curve generally shows a rapid increase in the leaf photosynthesis rate with increasing incident radiation when the radiation intensity is low and a much slower increase when the radiation intensity is high, i.e., photosynthesis becomes saturated at a high radiation intensity (Ogren, 1993; Perkins et al., 2006; Zheng et al., 2017). Such a light response curve suggests that the LUE in shaded leaves should be higher than that in sunlit leaves, because of the lower APAR absorbed by shaded leaves. As a result, the existing TL-LUE model with two prescribed LUEs for sunlit and shaded leaves has the following issues: (1) it does not capture the influence of radiation intensity variations on LUE for both sunlit and shaded leaves, and (2) the LUEs of sunlit and shaded leaves are independent of each other, while in reality they are related and their difference depends on the incoming radiation intensity, canopy structure, and other variables.

From the theoretical point of view, it is therefore necessary to develop a system in which sunlit and shaded leaf LUEs are determined mechanistically and cohesively. In some ecosystem models, such as BEPS (Chen et al., 1999; Liu et al., 1997) and BIOME-BGC (Hunt Jr and Running, 1992), a radiation scalar is used to scale stomatal conductance from its optimum value to reality. Since stomatal conductance and LUE are highly related, we attempt to use this scalar to improve the TL-LUE model. The main purpose of this study is to develop a modified TL-LUE model (RTL-LUE) with consideration of the influence of radiation on LUE using this radiation scalar and to demonstrate the improvements of RTL-LUE over TL-LUE in simulating GPP of terrestrial ecosystems derived from eddy covariance (EC) data for 12 vegetation types across the globe.

## 2. Data and methods

### *2.1 Data*

In this study, the FLUXNET2015 dataset is applied to the parameterization and validation of the RTL-LUE and TL-LUE models, including the records of GPP, incoming solar radiation, air temperature, and VPD. The FLUXNET2015 dataset ([www.fluxdata.org](www.fluxdata.org)) is the latest published flux observations from EC towers with a standard format, including the refined data from different regional networks for 212 sites worldwide (Pastorello et al., 2017; Pastorello et al., 2020). Similar to many previous studies, GPP partitioned by the DT method (termed "GPP_DT_VUT_REF") is employed as the true GPP, which is partitioned from NEE using a hyperbolic light response curve (Reichstein et al., 2005). The sites with more than one-year of valid half-hour observations during 2001−2015 are all selected, and there are in total 169 sites and 1191 site years covering 12 vegetation types in the International Geosphere-Biosphere Program (IGBP) classification system. The number of sites and site-years for each vegetation type are listed in Table 1, and the detailed information and references for each site can be found in Table S1. The half-hourly data are used because they can support the comparison of diurnal GPP variations.

Table 1. Number of EC sites and site years for different vegetation types in the IGBP classification system.

| Vegetation type [a] | DBF | DNF | EBF | ENF | MF | GRA | CRO | CSH | OSH | WET | SAV | WSA |
|---|---|---|---|---|---|---|---|---|---|---|---|---|
| Number of Site | 19 | 1 | 12 | 41 | 8 | 32 | 13 | 1 | 11 | 18 | 7 | 6 |
| Number of Site-years | 147 | 3 | 70 | 326 | 76 | 214 | 103 | 4 | 68 | 86 | 45 | 49 |

[a] DBF: deciduous broadleaf forest; DNF: deciduous needleleaf forest; ENF: evergreen needleleaf forest; EBF: evergreen broadleaf forest; MF: mixed forest; GRA: grass; CRO: crop; CSH: closed shrub; OSH: open shrub; WET: wetlands; SAV: savannas; WSA: woody savannas.

The Global LAnd Surface Satellite (GLASS) LAI product from 2000 to 2015 is used to drive the TL-LUE and RTL-LUE models, as the description of vegetation canopy structure and used to calculate APAR (Xiao et al., 2016). The dataset is provided at a spatial resolution of 1 km and temporal resolution of 8 days, which has been proved to have a satisfactory performance in GPP simulation (Xiao et al., 2017; Xie et al., 2019). In order to reduce the impacts of the residual cloud contamination and eliminate the unrealistically short-term fluctuations, the LAI time series data were smoothed with the Whittaker trend filtering method (Chu et al., 2021).

## 2.2 TL-LUE model

The TL-LUE model (He et al., 2013) stems from the MOD17 algorithm and improves the calculation of canopy APAR and GPP after separating the canopy into sunlit and shaded leaf groups according to the BEPS model (Chen et al., 1999). GPP of a plant canopy is calculated as follows:

$$GPP = \left( \varepsilon_{msu} \times APAR_{su} + \varepsilon_{msh} \times APAR_{sh} \right) \times f(VPD) \times g(T_a) \qquad (1)$$

where $\varepsilon_{msu}$ and $\varepsilon_{msh}$ are the maximum light use efficiency of sunlit and shaded leaves, respectively; $APAR_{su}$ and $APAR_{sh}$ are the incoming PAR absorbed by sunlit and shaded leaves

and calculated as follows:

$$APAR_{su} = (1-\alpha) \times \left[ \frac{PAR_{dir} \times \cos(\beta)}{\cos(\theta)} + \frac{PAR_{dif} - PAR_{dif,u}}{LAI} + C \right] \times LAI_{su} \quad (2)$$

$$APAR_{sh} = (1-\alpha) \times \left[ \frac{PAR_{dif} - PAR_{dif,u}}{LAI} + C \right] \times LAI_{sh} \quad (3)$$

where $\alpha$ is the canopy albedo that obtained related to vegetation types (Table. S2); $\beta$ is the mean leaf-sun angle and the value for a canopy with spherical leaf angle distribution is set as 60°; $\theta$ is the solar zenith angle; $PAR_{dif}$, $PAR_{dir}$, and $PAR_{dif,u}$ are the diffuse, direct components of incoming PAR, and the diffuse PAR under the canopy, respectively, which are empirically partitioned according to the clearness index ($R$) following Chen et al. (1999); $C$ represents the multiple scattering of direct radiation; $LAI_{su}$ and $LAI_{sh}$ are the LAI of sunlit and shaded leaves, separated according to the canopy LAI, clumping index ($\Omega$) and solar zenith angle according to Chen et al. (1999). The $LAI_{su}$ can be computed as follow, and the $LAI_{sh}$ is the residual of $LAI$ minus $LAI_{su}$.

$$LAI_{su} = 2 \times \cos(\theta) \times (1 - \exp(-0.5 \times \Omega \times \frac{LAI}{\cos(\theta)})) \quad (4)$$

In equation (1), $f(VPD)$ and $g(T_a)$ are the scalars of $VPD$ and minimum temperature, which are the same as it in the MOD17 model (Running et al., 2004), calculated as follows:

$$f(VPD) = \begin{cases} 0 & VPD \geq VPD_{max} \\ \frac{VPD_{max} - VPD}{VPD_{max} - VPD_{min}} & VPD_{max} < VPD < VPD_{max} \\ 1 & VPD \leq VPD_{min} \end{cases} \quad (5)$$

$$g(T_a) = \begin{cases} 1 & T_a \geq T_{max} \\ \frac{T_a - T_{min}}{T_{max} - T_{min}} & T_{min} < T_a < T_{max} \\ 0 & T_a \leq T_{min} \end{cases} \quad (6)$$

where $VPD$ is the daily average VPD at day-time and $T_a$ is the daily minimum temperature, which can be both calculated from the half-hour data; $VPD_{max}$, $VPD_{min}$, $T_{max}$, and $T_{min}$ are parameters specific to vegetation types (Table S2).

## 2.3 Description of the RTL-LUE model

Due to the fact that the nutrient and environmental conditions between sunlit and shaded leaves are very similar, the difference in LUE between these two groups of leaves should mainly attribute to the different light intensity. According to the BEPS and BIOME-BGC models, the scalar of photosynthetic photon flux density (PPFD) can be used to quantify the response of stomatal conductance to light intensity (Chen et al., 1999; Hunt Jr and Running, 1992; Liu et al., 1997). After considering the scalar of PPFD for the photosynthesis of sunlit and shaded leaves, a TL-LUE model with radiation constraint on LUE (RTL-LUE) is developed, and the calculation of GPP can be modified as follows:

$$GPP = \varepsilon_{max}^* \times \left( f(PPFD_{su}) \times APAR_{su} + f(PPFD_{sh}) \times APAR_{sh} \right) \times f(VPD) \times f(T_a) \quad (7)$$

where $\varepsilon_{max}^*$ is the maximum LUE of all the leaves within the canopy; $APAR_{su}$, $APAR_{sh}$ $f(VPD)$ and $g(T_a)$ are calculated in the same way as in TL-LUE and the MOD17 model; $f(PPFD_{su})$ and $f(PPFD_{sh})$ are the radiation scalars for sunlit and shaded leaves, respectively, which are calculated by the PPFD of the two groups of leaves, using the same reciprocal function as follows:

$$f(PPFD) = \frac{b}{a \times PPFD + b} \quad (8)$$

where $a$ and $b$ are the coefficients in the relationship between light intensity and LUE. $b$ is set as a constant with the value of 1 mol m$^{-2}$ hh$^{-1}$, and only the parameter $a$ is used to control the response of LUE to PPFD and needs to be parameterized depending on different vegetation types. $PPFD$ (mol m$^{-2}$ hh$^{-1}$) for sunlit and shaded leaves can be obtained from the absorbed PAR ($PAR_{su}$ or $PAR_{sh}$) multiplied by a constant PAR-energy ratio of 4.55 mol/MJ (Chen et al., 1999).

## 2.4 Model Parameterization and evaluation

### 2.4.1 Model Parameterization

Similar to previous studies, the parameters $VPD_{max}$, $VPD_{min}$, $T_{max}$, $T_{min}$, $\Omega$ and $\alpha$ are empirically set both in TL-LUE and RTL-LUE models according to previous studies (He et al., 2013; Zhou et al., 2016), as shown in Table S2. Two parameters need to be further optimized in both the RTL-LUE model ($\varepsilon_{max}$ and $a$) and the TL-LUE model ($\varepsilon_{msu}$ and $\varepsilon_{msh}$). These four parameters were all optimized using the randomly selected one-year data for each site (Table S1), and the mean values of the sites with the same vegetation type are obtained. The shuffled complex evolution-University of Arizona method was employed to implement the optimization (Duan et al., 1992; Zhou et al., 2016), which evaluates the model performance with the agreement index (*d*):

$$d = 1 - \frac{\sum_{i=1}^{N}(P_i - O_i)^2}{\sum_{i=1}^{N}(|P_i - \overline{O}| + |O_i - \overline{O}|)^2} \quad (9)$$

where *N* is the total simulated experiment data point; $P_i$ and $O_i$ represent the predicted GPP from the TL-LUE or RTL-LUE models and EC GPP, respectively; and $\overline{O}$ and $\overline{P}$ are the mean values of observations and predictions for all experimental data points.

### 2.4.2 Model evaluation

The accuracies of the GPP simulations from the two models were assessed against half-hourly, composited 8-day, and composited yearly EC GPP. The diurnal and seasonal variations of GPP were qualitatively compared, and a simple ratio of EC GPP and model simulations were used to quantitatively assess the fluctuation of the difference between modeled and observed GPP. In addition, three indexes were used for accuracy evaluation, including the coefficient of determination

($R^2$), the root-mean-square error (RMSE), and the mean predictive error (bias):

$$R^2 = \left( \frac{\sum_{i=1}^{N}(P_i - \overline{O})(O_i - \overline{O})}{\sqrt{\sum_{i=1}^{N}(P_i - \overline{P})^2} \sqrt{\sum_{i=1}^{N}(O_i - \overline{O})^2}} \right)^2 \quad (10)$$

$$RMSE = \sqrt{\frac{\sum_{i=1}^{N}(P_i - O_i)^2}{N}} \quad (11)$$

$$bias = \frac{\sum_{i=1}^{N}(O_i - P_i)}{N} \quad (12)$$

## 3. Results

In this section, the performances in GPP estimation from the TL-LUE and RTL-LUE models are evaluated against the EC GPP. The results at half-hourly, 8-day, and yearly scales are all compared first, both for all or individual vegetation types. Then, the diurnal and seasonal variations of GPP simulations from the two models are further evaluated, based on the half-hourly data and the composited 8-day data, respectively.

*3.1 Validation of the GPP simulations*

**3.1.1 Half-hourly results**

In this study, GPP was estimated based on the half-hourly EC observation datasets, and the accuracy of simulated results was first evaluated at the half-hour scale, and further assessed after they were composited into the 8-day and yearly GPP in the following two sections. As shown in Fig. 1, $R^2$ and RMSE for GPP simulations from the TL-LUE and RTL-LUE models were calculated at 169 EC sites against the EC GPP, and compared in a scatterplot. It can be observed that the RTL-LUE model outperforms the TL-LUE model at almost all sites, with a higher $R^2$ and lower RMSE.

There are in total 160 out of 169 sites showing a reduced RMSE after considering the radiation scalar in the RTL-LUE model, and the number is 144 for sites with a higher $R^2$. The largest improvement in $R^2$ is 0.097, and the largest reduction in RMSE is 0.15 g C m$^{-2}$ hh$^{-1}$. There are 102 sites showing an improvement in $R^2$ more than 0.02, and 111 sites showing a reduction in RMSE more than 0.02 g C m$^{-2}$ hh$^{-1}$. The improvements in $R^2$ are even larger and more stable at the sites with higher $R^2$ values, as shown in Fig. 1 (a). For example, there are in total 56 sites with $R^2$ greater than 0.7, and 54 out of them show an improved $R^2$ with a mean increase of 0.04. On the contrary, the reductions in RMSE are more remarkable at the sites with higher RMSE, as shown in Fig. 3 (b). Among the 70 sites with RMSE higher than 0.15, 69 of them show a lower RMSE in the RTL-LUE model with a mean reduction of 0.068 g C m$^{-2}$ hh$^{-1}$.

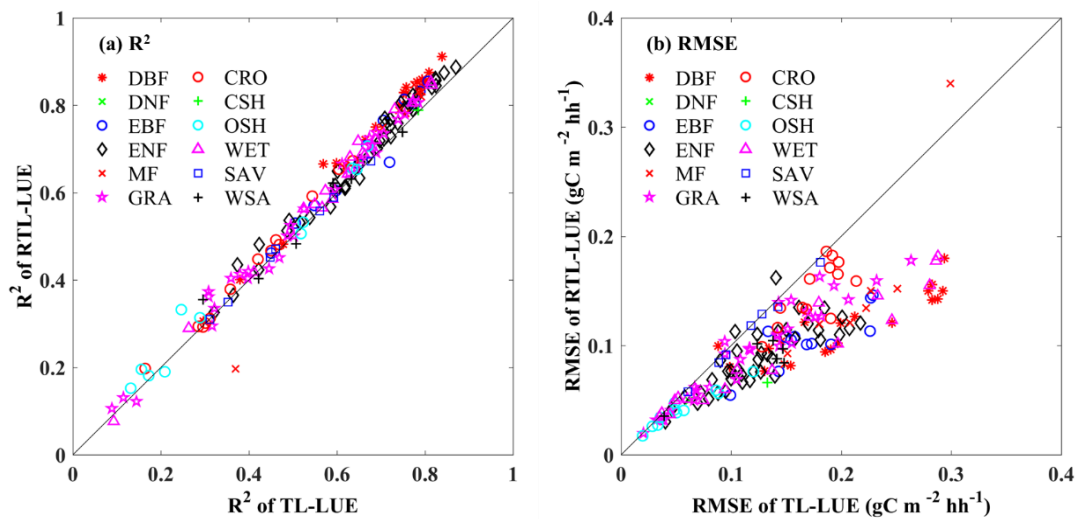

**Figure 1**. Comparison of the accuracy of GPP simulations against tower measurements (EC GPP) for the original TL-LUE model and the RTL-LUE model at 169 EC observation sites: (a) R2, (b) RMSE.

The three indexes of GPP simulations from the TL-LUE and RTL-LUE models were also compared among the 12 different vegetation types, by calculating the mean value for all sites with the same vegetation type. As shown in Table. 3, although the GPP simulated by the TL-LUE model agrees well with the EC data, the RTL-LUE model further improves the simulation of GPP in all of the 12 vegetation types. Overall, the TL-LUE model shows mean values of 0.58 and 0.14 g C m$^{-2}$ hh$^{-1}$ for R$^2$ and RMSE, respectively, and the corresponding values are 0.61 and 0.10 g C m$^{-2}$ hh$^{-1}$ for the RTL-LUE model with an improvement in R$^2$ by 0.03 and a reduction in RMSE by 0.04 g C m$^{-2}$ hh$^{-1}$. The rank of vegetation types based on the value of R$^2$, RMSE, or bias are all the same. The value of R$^2$ ranges from 0.38 for open shrubs to 0.78 for closed shrubs in the results of the TL-LUE model, and the range is from 0.40 to 0.82 for the RTL-LUE model. The RTL-LUE model consistently performs much better in the five types of forest sites, and a considerable improvement is found in the deciduous broadleaf forest sites, with a difference of 0.06, 0.09 g C m$^{-2}$ hh$^{-1}$, and 0.06 g C m$^{-2}$ hh$^{-1}$ for R$^2$, RMSE, and bias, respectively. However, almost no improvements can be observed at the sites covered by savannas, only with a limited reduction in RMSE and bias of about 0.01 g C m$^{-2}$ hh$^{-1}$.

**Table 2.** Statistic of the GPP simulation accuracy for different vegetation types.

| Vegetation type [a] | TL-LUE | | | RTL-LUE | | | Difference [b] | | |
|---|---|---|---|---|---|---|---|---|---|
| | R$^2$ | RMSE [c] | bias [c] | R$^2$ | RMSE [c] | bias [c] | R$^2$ | RMSE [c] | bias [c] |
| **DBF** | 0.68 | 0.20 | 0.11 | 0.74 | 0.12 | 0.04 | 0.06 | -0.09 | -0.06 |
| **DNF** | 0.65 | 0.06 | 0.02 | 0.67 | 0.05 | 0.02 | 0.02 | -0.01 | -0.01 |
| **EBF** | 0.66 | 0.17 | 0.07 | 0.69 | 0.11 | 0.02 | 0.03 | -0.07 | -0.05 |
| **ENF** | 0.66 | 0.12 | 0.03 | 0.68 | 0.09 | 0.00 | 0.02 | -0.04 | -0.03 |
| **MF** | 0.71 | 0.21 | 0.05 | 0.73 | 0.15 | 0.01 | 0.02 | -0.06 | -0.04 |
| **GRA** | 0.51 | 0.12 | 0.04 | 0.53 | 0.10 | 0.02 | 0.02 | -0.03 | -0.02 |
| **CRO** | 0.41 | 0.18 | 0.06 | 0.43 | 0.15 | 0.03 | 0.02 | -0.03 | -0.03 |
| **CSH** | 0.78 | 0.13 | 0.08 | 0.82 | 0.07 | 0.01 | 0.04 | -0.07 | -0.07 |
| **OSH** | 0.38 | 0.06 | 0.02 | 0.40 | 0.04 | 0.01 | 0.02 | -0.01 | -0.01 |
| **WET** | 0.59 | 0.14 | 0.07 | 0.62 | 0.09 | 0.03 | 0.03 | -0.05 | -0.04 |
| **SAV** | 0.49 | 0.12 | 0.02 | 0.49 | 0.11 | 0.01 | 0.00 | -0.01 | -0.01 |
| **WSA** | 0.53 | 0.12 | 0.07 | 0.54 | 0.09 | 0.03 | 0.01 | -0.04 | -0.04 |
| **All** | 0.58 | 0.14 | 0.05 | 0.61 | 0.10 | 0.02 | 0.03 | -0.04 | -0.03 |

[a] The abbreviation of different vegetation types are the same as it in Table 1; [b] Difference means the R$^2$, RMSE,

and bias of RTL-LUE model minus the corresponding value of RL-LUE model; [c] the units of RMSE and bias are g C m$^{-2}$ hh$^{-1}$.

### 3.1.2 8-day results

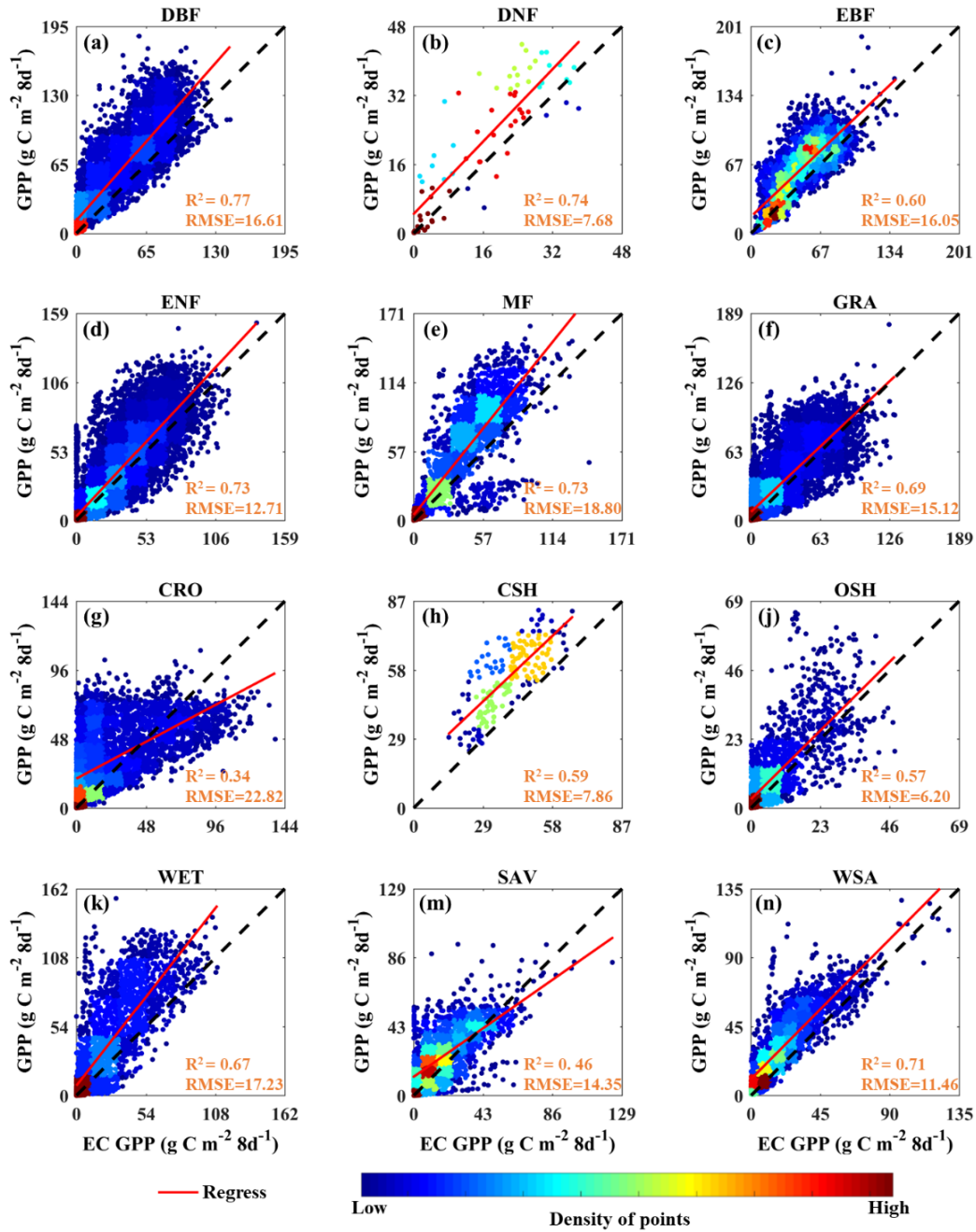

**Figure 2.** Validation of GPP simulations from the TL-LUE model after composited into 8-day totals.

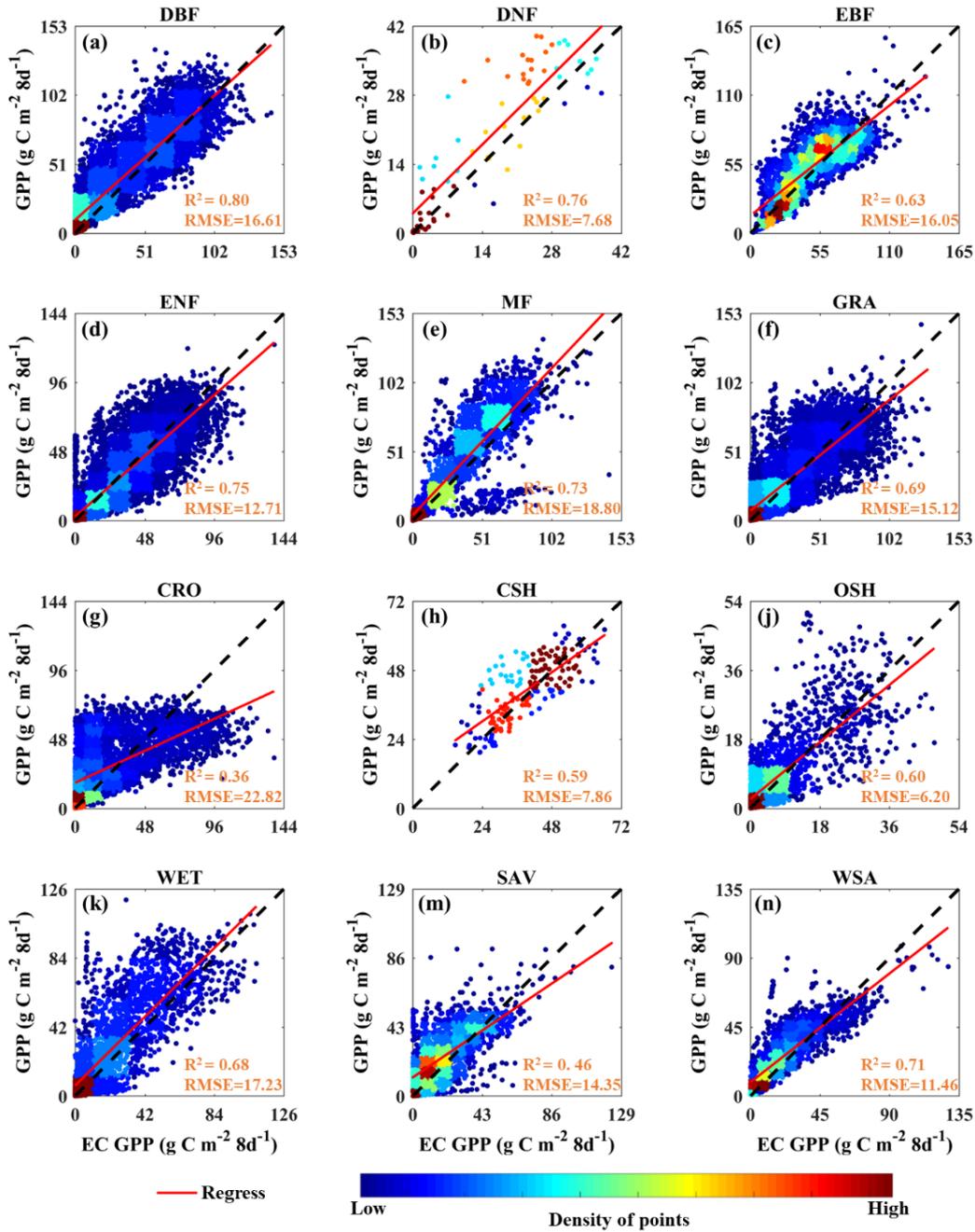

**Figure 3.** Validation of GPP simulations from the RTL-LUE model after composited into 8-day totals.

In order to further evaluate the accuracy of GPP simulated by the TL-LUE and RTL-LUE models, the half-hourly data were composited using the 8-day sum to match the time interval of LAI data. The density scatterplot of all the data points for the 12 individual vegetation types are provided in Fig.3 (TL-LUE) and Fig.4 (RTL-LUE), respectively. It can be observed that the overall accuracies of both the two models are improved after the 8-day composition, with a mean $R^2$ of 0.64 and 0.65 for the GPP simulations from the TL-LUE and RTL-LUE models, and the RMSE is 19.01 g C m$^{-2}$ 8d$^{-1}$ and 13.91 g C m$^{-2}$ 8d$^{-1}$, respectively. The $R^2$ of grass, woody savannas, and open shrubs sites even show an improvement in $R^2$ greater than 0.18. However, the accuracies of GPP simulations in some vegetation types decrease compared to the half-hourly data, and typically the closed shrubs sites show a decline in $R^2$ by −0.19. For the 8-day data, the most significant improvement of the RTL-LUE model over the TL-LUE model is found at open shrub sites with an increase of 0.03 for $R^2$ and a decrease of 1.95 g C m$^{-2}$ 8d$^{-1}$ for RMSE. The best accuracy can be found in the deciduous broadleaf forest sites with an $R^2$ of 0.77 and 0.80 for the TL-LUE and RTL-LUE models, respectively. The worst performance is found at crop sites with an $R^2$ of 0.34 and 0.36 for the two models, respectively. GPP of crops are not well simulated in both two models, with an overestimation in low EC GPP points and underestimation in high EC GPP points. Similar simulation results for crops were reported in previous studies using the BEPS model and the MOD17 model, and failure to account for the impacts on irrigation and fertilization in crops was considered to be the main reason for the low performance of these models (Zhang et al., 2012).

### 3.1.3 Yearly results

In order to conduct an overall assessment of GPP simulated by the TL-LUE and RTL-LUE models, half-hourly data were further summed into yearly data. The density scatterplots of simulated GPP against the EC GPP of all site-years are shown in Fig. 4. Although GPP simulated by the TL-LUE model already shows good agreement with EC GPP, the results from the RTL-LUE model further confirm its superiority, with improvements of $R^2$ (+0.02), RMSE (−165.28 g C m$^{-2}$ year$^{-1}$), and bias (−202.59 g C m$^{-2}$ year$^{-1}$). Furthermore, it can be observed that the TL-LUE model tends to overestimate GPP, especially in the points with high GPP values. This phenomenon is consistent in Fig.2. However, this problem is relieved by the RTL-LUE model after considering the radiation scalar, which could suppress the high value of GPP when radiation was high.

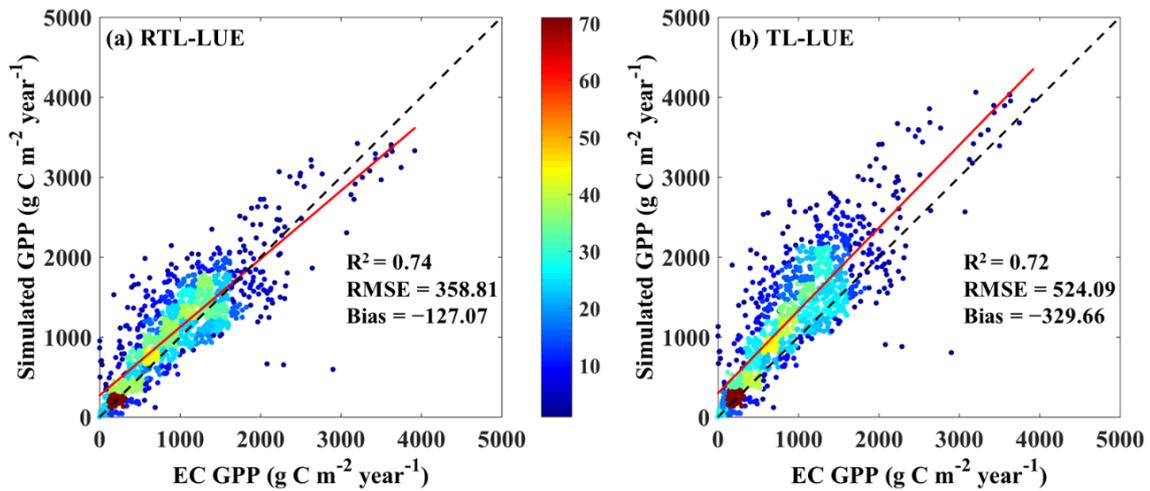

**Figure 4.** The relation between the yearly EC GPP and the simulated results from (a) the RTL-LUE model and (b) the TL-LUE model.

In general, from the analysis above, it can be summarized that the RTL-LUE model can improve the accuracy in GPP simulation after using the radiation scalar, at half-hourly, 8-day, and yearly time scales. For all or individual vegetation types, the GPP simulated by the RTL-LUE model shows increased $R^2$ and reduced RMSE and bias relative to the TL-LUE model in comparison with EC observations.

## 3.2 Diurnal variation of GPP

**Table 3.** Statistics of the diurnal variation consistency between half-hourly GPP simulations and EC GPP over all site-days.

| Vegetation types [a] | TL-LUE | | | RTL-LUE | | | Percentage [b] (%) | | |
|---|---|---|---|---|---|---|---|---|---|
| | $R^2$ | RMSE[c] | R_CV[d] (%) | $R^2$ | RMSE[c] | R_CV[d] (%) | $R^2$ | RMSE[c] | R_CV[d] |
| DBF | 0.91 | 0.16 | 33.24 | 0.95 | 0.09 | 17.83 | 83.95 | 78.17 | 90.75 |
| DNF | 0.91 | 0.05 | 33.58 | 0.93 | 0.04 | 22.67 | 68.78 | 80.41 | 89.18 |
| EBF | 0.92 | 0.14 | 31.69 | 0.96 | 0.09 | 18.27 | 93.81 | 80.73 | 97.63 |
| ENF | 0.91 | 0.09 | 32.44 | 0.95 | 0.06 | 21.53 | 94.58 | 62.93 | 97.41 |
| MF | 0.90 | 0.16 | 29.00 | 0.91 | 0.11 | 20.00 | 91.31 | 79.44 | 94.15 |
| GRA | 0.85 | 0.09 | 39.37 | 0.88 | 0.07 | 28.57 | 81.29 | 71.28 | 91.35 |
| CRO | 0.90 | 0.12 | 33.02 | 0.94 | 0.10 | 19.17 | 84.88 | 68.25 | 91.54 |
| CSH | 0.94 | 0.12 | 26.08 | 0.95 | 0.06 | 22.45 | 95.75 | 86.28 | 99.38 |
| OSH | 0.77 | 0.04 | 53.32 | 0.82 | 0.03 | 39.16 | 84.10 | 74.59 | 95.04 |
| WET | 0.92 | 0.11 | 29.73 | 0.95 | 0.07 | 18.63 | 84.17 | 77.17 | 91.30 |
| SAV | 0.78 | 0.09 | 41.36 | 0.78 | 0.08 | 40.51 | 50.78 | 73.95 | 79.83 |
| WSA | 0.79 | 0.10 | 41.52 | 0.79 | 0.07 | 31.59 | 58.23 | 84.06 | 87.36 |
| All | 0.88 | 0.10 | 35.40 | 0.91 | 0.07 | 23.95 | 83.55 | 76.49 | 92.09 |

[a] The abbreviation of different vegetation types are the same as it in Table 1; [b] Percentage means the percentage of site-days that the diurnal variation of GPP simulated by RTL-LUE model was better than TL-LUE model, i.e., with a higher $R^2$, lower RMSE, and lower R_CV; [c] unit of RMSE is g C m$^{-2}$ hh$^{-1}$; [d] R_CV means the Coefficients of Variation (CV) of the ratio (EC GPP/simulated GPP), and represents the diurnal fluctuation of the difference between GPP simulations and EC observations.

Since GPP was calculated based on the half-hourly dataset, the diurnal variation of GPP simulated by the TL-LUE and RTL-LUE models can be compared against EC data. In order to validate the diurnal variation of GPP, the Coefficients of Variation (CV) of the Ratio (R_CV) between EC GPP and simulated GPP (Ratio=EC GPP/simulated GPP) in each site-days were calculated. Since the Ratio represents the relative difference between the simulated GPP and EC GPP, the R_CV, which denotes the variation of Ratio in each day, can certainly assess the consistency of the diurnal variation between EC GPP and GPP simulations. If the diurnal variation of simulated GPP is entirely the same as the EC GPP, the R_CV should be zero. As a result, the smaller the R_CV, the better the diurnal variation of the simulated GPP is. In addition, the $R^2$ and RMSE between the simulated GPP and EC GPP in all the 333,919 site-days were also calculated, and the mean values of them for 12 vegetation types are displayed in Table. 4. It can be observed that the mean R_CV of GPP simulations from the RTL-LUE model is obviously lower than that from the TL-LUE model, either in all or individual vegetation types. Among all the 333,919 site-days, 92.9% of them show a lower R_CV in the GPP simulations from the RTL-LUE model compared to the TL-LUE model simulations, while the ratio is 83.4% and 76.5% for a higher $R^2$ and lower RMSE. The overall improvement of R_CV reaches 11.5%, which is almost one-third of the R_CV for the GPP estimations from the TL-LUE model. The most remarkable improvement can be found at the sites covered by evergreen broadleaf forest and open shrubs, with differences greater than 0.14 in R-CV and 0.04 in $R^2$. The diurnal variation of GPP at savannas sites are similar between the two models, with almost the same $R^2$ and very small differences in RMSE and R_CV.

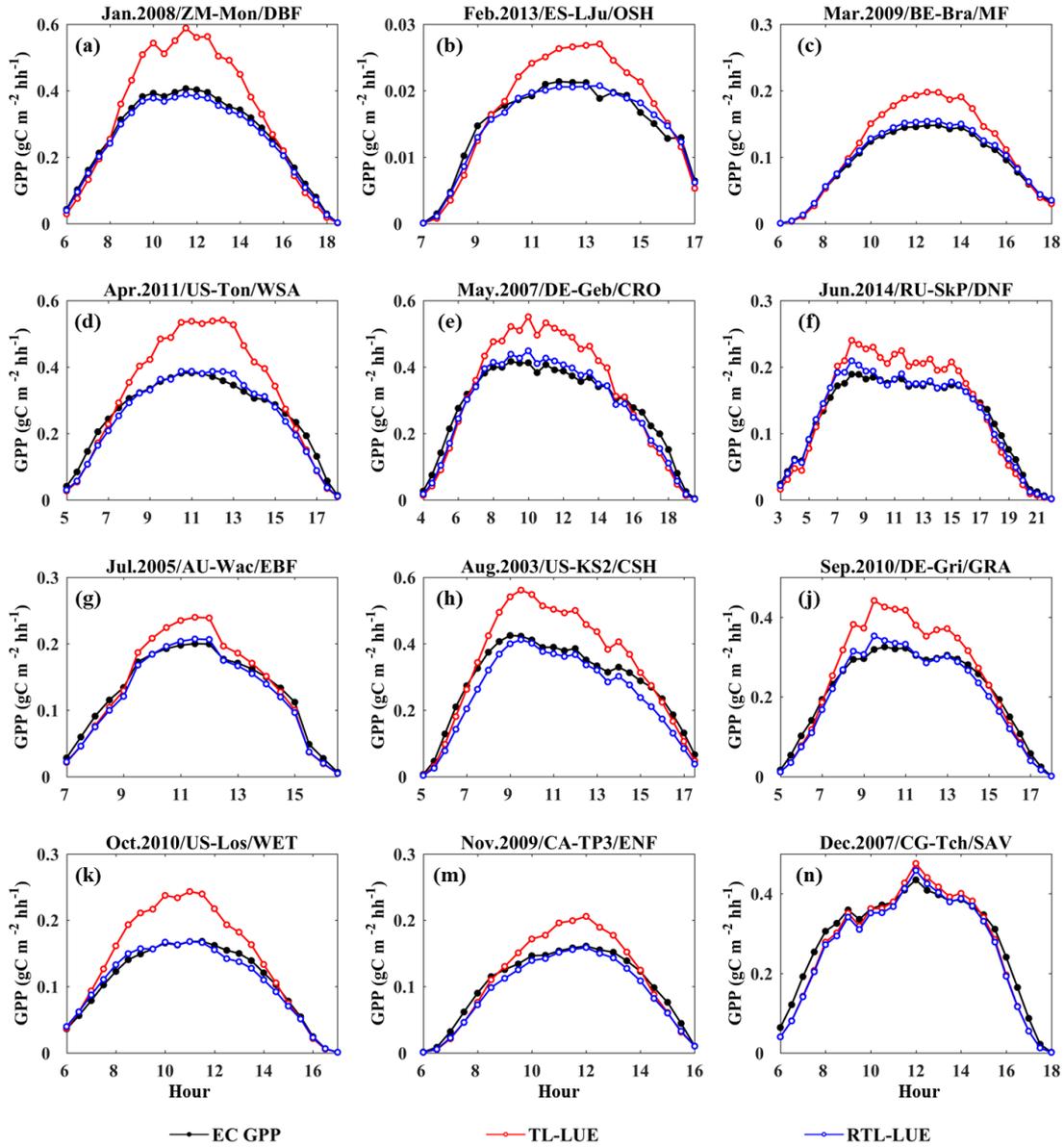

**Figure 5**. Diurnal variation of EC GPP and GPP simulated by the TL-LUE and RTL-LUE models in different months and different vegetation types. The title of each subplot is named as "month. year/site name/vegetation types".

The diurnal variation of GPP estimations was further qualitatively compared against the EC observations, by plotting representative diurnal variation curves from 12 different months and 12 different vegetation types. As shown in Fig. 5, monthly mean values for each half-hour were calculated and plotted. The monthly means were used in order to minimize the fluctuations in a single day and to enhance the dependability of the comparisons. For all the months and vegetation types, both the GPP simulations from the two models and EC GPP consistently reaches the peak at noon with the highest radiation. However, in almost all the 12 plots, the TL-LUE model tends to overestimate GPP at the peak, while the GPP simulated by the RTL-LUE model agrees well with EC GPP in this condition. The improvement in simulating the peak GPP values with the RTL-LUE model can mainly attribute to the use of the radiation scalar in the model. Because no radiation scalar is used in the TL-LEU model, the LUEs of both sunlit and shaded leaves do not vary with radiation, causing underestimation of GPP at low PAR but overestimation at high PAR, even though the LUEs are optimized using the observed GPP to capture their mean values. The use of the radiation scalar avoids this issue, and this is the main improvement of RTL-LUE over TL-LUE. While in the morning and dusk, when the radiation scalar does not play an important role because radiation is generally low, both the GPP simulated by TL-LUE and RTL-LUE models agree well with the EC GPP. For the sites covered by the savannas, the diurnal variation of GPP simulated by the two models are almost the same, which is mostly due to the low value of *a* used (0.20, the lowest among 12 vegetation types), so the variation of GPP will not be sensitive to the intensity of radiation.

## 3.3 Seasonal variation of GPP

The seasonal variation is also critical when evaluating the reliability of GPP simulations. Similar to the assessment of diurnal variation, the composited 8-day data were employed to calculate $R^2$, RMSE, and R_CV between the simulated GPP and EC data for all site-years. The 8-day data were used to eliminate the diurnal variation and only assess the seasonal variation. The mean values of the three indexes for 12 vegetation types are summarized in Table 4. There are some differences between the results in Section 3.1.2, where the indexes are calculated for each site, but the evaluation here is made for individual site-years. The conclusions are consistent with those for the diurnal variation. The seasonal variation of GPP simulated by the RTL-LUE model agrees better with EC GPP than that by the TL-LUE model, either in all or individual vegetation types. Overall, about 92.3% of the 1191 site-years show a declined R_CV for the GPP simulations from the RTL-LUE model comparing to the TL-LUE model results, and the ratio is 87.5% and 81.3% for a higher $R^2$ and lower RMSE for RTL-LUE. The quantitative improvements in seasonal variation are not as significant as those in the diurnal variation. The most considerable decrease in R_CV is 4.5% in ENF, and the greatest improvement in $R^2$ is 0.02 and the greatest reduction in RMSE is 9.86 g C $m^{-2}$ $8d^{-1}$ in DNF and CSH. Although RMSE and R_CV decrease in all 12 vegetation types when TL-LUE is replaced with RTL-LUE, $R^2$ is almost the same for SAV, CSH, and ENF.

**Table 4.** Statistics of the seasonal variation consistency between 8-day GPP simulations and EC GPP over all site-years.

| Vegetation type [a] | TL-LUE | | | RTL-LUE | | | Percentage [a] (%) | | |
|---|---|---|---|---|---|---|---|---|---|
| | $R^2$ | RMSE [b] | R_CV [c] (%) | $R^2$ | RMSE [b] | R_CV [c] (%) | $R^2$ | RMSE [b] | R_CV [c] |
| DBF | 0.83 | 24.66 | 52.80 | 0.84 | 15.67 | 50.00 | 95.68 | 87.05 | 81.29 |
| DNF | 0.83 | 8.34 | 44.94 | 0.85 | 7.16 | 41.69 | 100.00 | 100.00 | 100.00 |
| EBF | 0.48 | 22.02 | 37.47 | 0.49 | 15.28 | 31.91 | 66.67 | 91.30 | 98.55 |
| ENF | 0.84 | 14.41 | 54.83 | 0.84 | 11.28 | 49.81 | 64.09 | 69.97 | 99.07 |
| MF | 0.90 | 24.48 | 46.02 | 0.91 | 16.71 | 40.95 | 87.32 | 95.77 | 95.77 |
| GRA | 0.64 | 14.68 | 62.65 | 0.65 | 12.57 | 58.92 | 76.53 | 75.59 | 94.37 |
| CRO | 0.41 | 24.68 | 74.77 | 0.42 | 21.97 | 72.01 | 90.29 | 90.29 | 92.23 |
| CSH | 0.63 | 17.69 | 25.67 | 0.63 | 7.83 | 24.90 | 75.00 | 75.00 | 75.00 |
| OSH | 0.53 | 6.94 | 73.82 | 0.54 | 5.64 | 68.99 | 69.12 | 79.41 | 97.06 |
| WET | 0.82 | 20.47 | 57.10 | 0.83 | 14.16 | 54.53 | 80.49 | 71.95 | 81.71 |
| SAV | 0.63 | 13.05 | 48.07 | 0.63 | 12.71 | 47.48 | 61.36 | 93.18 | 79.55 |
| WSA | 0.57 | 14.56 | 48.72 | 0.58 | 10.08 | 44.20 | 83.67 | 95.92 | 93.88 |
| All | 0.71 | 17.59 | 52.24 | 0.72 | 13.40 | 48.78 | 81.27 | 87.54 | 92.79 |

[a] The abbreviation of different vegetation types are the same as it in Table 1; [b] Percentage means the percentage of

site-days that the diurnal variation of GPP simulated by RTL-LUE model was better than TL-LUE model, i.e., with a higher $R^2$, lower RMSE, and lower R_CV; [c] unit of RMSE is g C m$^{-2}$ 8d$^{-1}$; [d] R_CV means the CV of the ratio (EC GPP/simulated GPP), and represents the seasonal fluctuation of the difference between GPP simulations and EC observations.

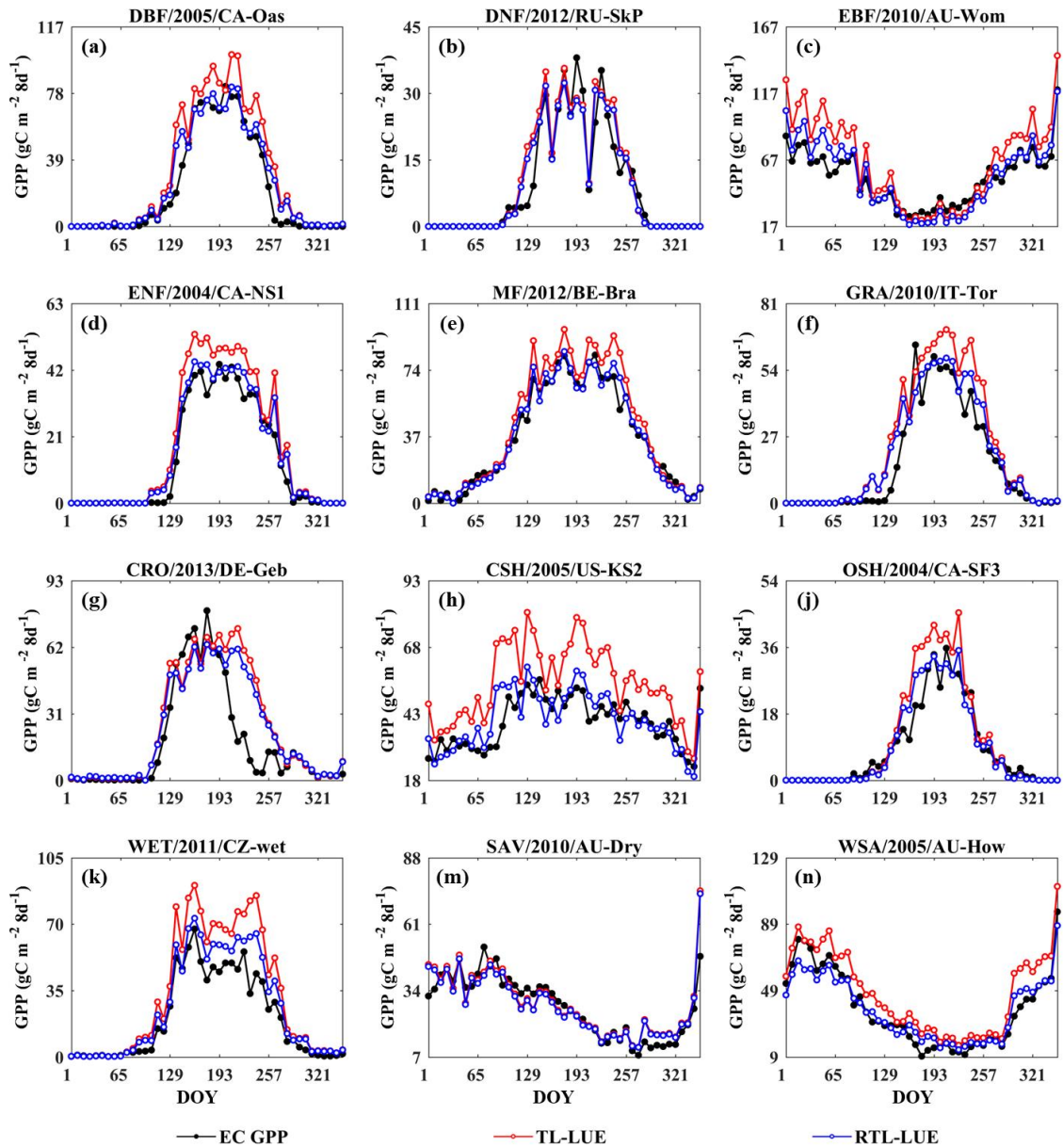

**Figure 6.** Seasonal variations of EC GPP and GPP simulated by the TL-LUE and RTL-LUE models for different vegetation types. The title of each subplot is named as "vegetation types/site year/site name", DOY means the day of year.

The seasonal variation curves of some representative site years for the 12 vegetation types were also selected for qualitative assessment, as displayed in Fig. 6. The seasonal variation of GPP simulated by both models show very similar seasonal variation with EC GPP in almost all vegetation types, except for crops. The apparent decline of EC GPP during the DOY (day of the year) from 193 to 257, which may be caused by crop harvest, cannot be well simulated by both models. It may be mostly caused by the uncertainties in LAI data, which cannot sufficiently capture crop growth stages. The unsatisfactory results for the evaluation of crops in Section 3.2.2 may also be caused by the same reason, as the low values in EC GPP during the harvest period are all overestimated in Fig. 2 and 3 (g). Furthermore, the overestimation of GPP by the TL-LUE model can also be found in the growing season, such as the vegetation types of DBF, EBF, ENF GRA, and CSH. The seasonal variation of GPP simulated by the two models is almost the same for the DNF and SAV sites, and both of them agreed well with the EC GPP.

In general, the TL-LUE GPP initially agrees well with the EC observations in most vegetation types, but the RTL-LUE model can further improve the consistency with observations. Although the seasonal variations of GPP simulated by the RTL-LUE model are better than those by the TL-LUE model in most site years, the improvements are not as significant as it in the diurnal variation, either in the quantitative and qualitative analysis.

## 4. Discussion

In this section, the relationship between PAR and canopy LUE is first discussed to exam the reasonability of the radiation scalar used in the RTL-LUE model. Then the parameterized different maximum LUEs in the two models are compared before analyzing the sensitivity of GPP estimation and canopy LUE to the variation of PAR. Afterward, the reason for the overestimation of GPP in the TL-LUE model is further explored. The improvements and limitations of the RTL-LUE model are summarized at last.

## *4.1 Relationship between PAR and canopy LUE in site observations*

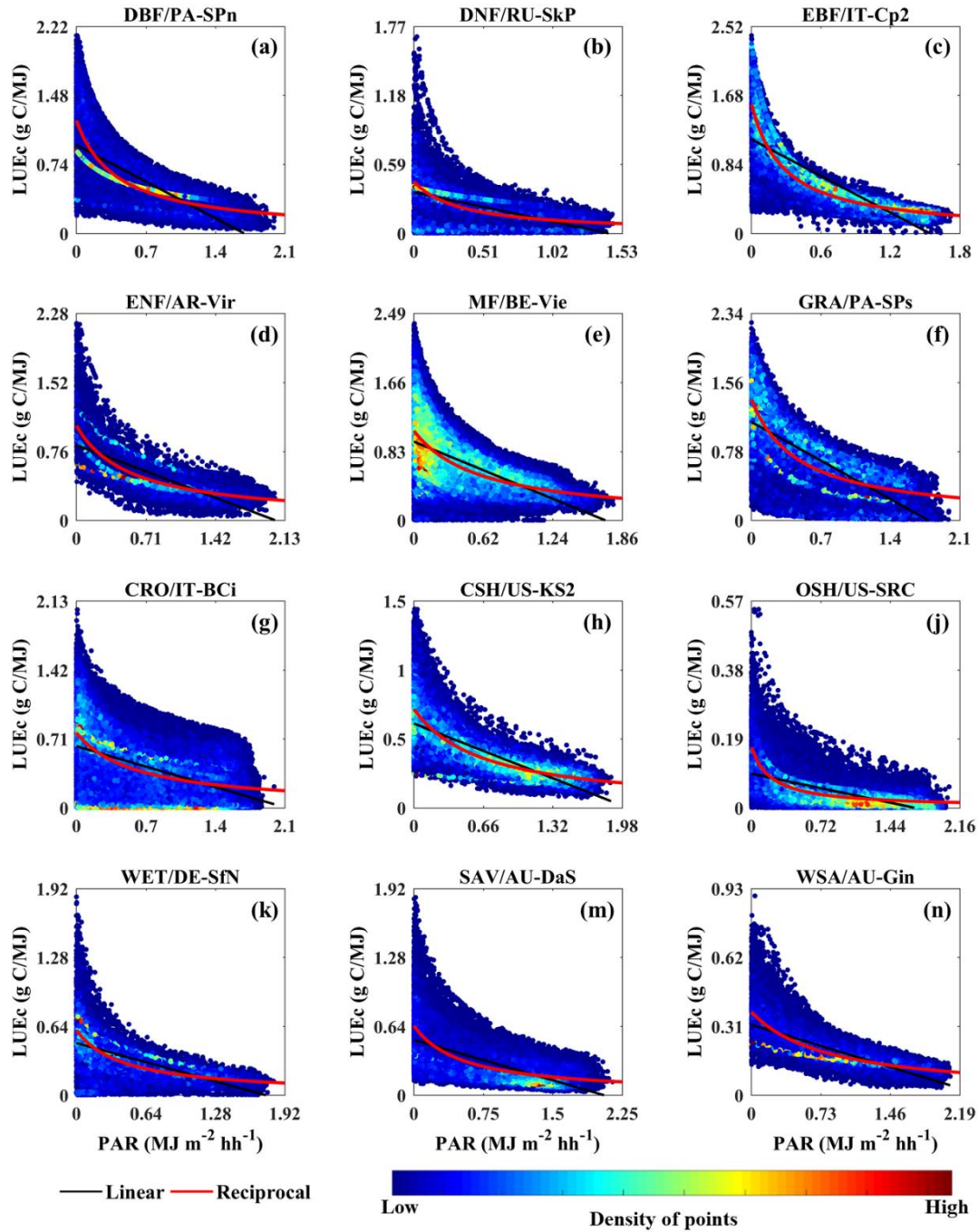

**Figure 7.** Comparison of the linear and reciprocal relationship between canopy LUE(LUEc=GPP/PAR) and PAR using site observations. The title of each subplot is named as "vegetation types/site name".

In order to the exam reasonability of adopting the scalar of radiation in the GPP simulation, it is necessary to verify the relationship between LUE and PAR based on the site observations. Since there are no LUE measurements at the EC sites, the canopy LUE was approximated by the ratio of the observed GPP and PAR (LUEc=GPP/PAR) (He et al., 2013). LAI or APAR was not employed here to avoid errors induced by remote sensing data, so the analysis based on measurements only can capture the true relationship between LUE and PAR. As shown in Fig. 7, the density scatterplot of LUEc and PAR at specific sites among 12 vegetation types are demonstrated, and the linear and reciprocal fitting lines are also included. The LUEc decreases with the increase in PAR in all vegetation types. However, the decrease is not linear, with the rate of decrease higher at the lower PAR values and gradually becoming slower with increasing PAR. As a result, it can be concluded that the reciprocal function can better fit the relationship between LUEc and PAR, and is suitable to represent the scalar of radiation on GPP estimation.

Furthermore, the quantitative assessments were also conducted by calculating the mean values of $R^2$ and RMSE for the linear and reciprocal regressions for the 12 vegetation types. The index of bias was excluded because it is too small to compare. As listed in Table 5, the reciprocal function can fit the relationship between LUEc and PAR much better than the linear regression, with higher $R^2$ and lower RMSE. Overall, the reciprocal function outperforms the linear relationship in 98.8% of the 169 sites. Only two sites covered by EDF and GRA (site names are AU-Ync and FR-Fon) show an $R^2$ lower than 0.05 for both two functions. The improvements of $R^2$ ranges from 0.01 in crops to 0.07 in EBF. Both methods perform the worst in crops, only with an $R^2$ lower than 0.1 and very high RMSE around 0.34 g C MJ$^{-2}$. It may also be the reason for the unsatisfactory result in GPP simulations by both TL-LUE and RTL-LUE models. Even in this condition, the reciprocal regression is better than the linear regression at all crop sites. In general, it can be concluded that the canopy LUE (approximated by LUEc=GPP/PAR) shows a reciprocal relationship with PAR, so it is reasonable to use the radiation scalar in the RTL-LUE model.

**Table 5.** Comparing the linear and reciprocal relationships between LUEc (GPP/PAR) and PAR based on site observations for different vegetation types.

| Vegetation type [a] | Linear | | Reciprocal | | Difference | | Percentage [b] (%) | |
|---|---|---|---|---|---|---|---|---|
| | $R^2$ | RMSE [c] | $R^2$ | RMSE [c] | $R^2$ | RMSE [b] | $R^2$ | RMSE [c] |
| DBF | 0.13 | 0.35 | 0.15 | 0.34 | 0.02 | -0.01 | 94.74 | 94.74 |
| DNF | 0.23 | 0.17 | 0.28 | 0.16 | 0.05 | -0.01 | 1 | 1 |
| EBF | 0.57 | 0.18 | 0.64 | 0.16 | 0.07 | -0.02 | 1 | 1 |
| ENF | 0.28 | 0.23 | 0.33 | 0.22 | 0.05 | -0.01 | 1 | 1 |
| MF | 0.22 | 0.28 | 0.24 | 0.27 | 0.02 | -0.01 | 1 | 1 |
| GRA | 0.19 | 0.23 | 0.23 | 0.22 | 0.04 | -0.01 | 96.88 | 96.88 |
| CRO | 0.07 | 0.34 | 0.08 | 0.34 | 0.01 | -0.00 | 1 | 1 |
| CSH | 0.53 | 0.14 | 0.58 | 0.13 | 0.05 | -0.01 | 1 | 1 |
| OSH | 0.22 | 0.11 | 0.29 | 0.10 | 0.06 | -0.01 | 1 | 1 |
| WET | 0.15 | 0.23 | 0.17 | 0.22 | 0.02 | -0.01 | 1 | 1 |
| SAV | 0.25 | 0.19 | 0.28 | 0.18 | 0.03 | -0.01 | 1 | 1 |
| WSA | 0.29 | 0.15 | 0.34 | 0.14 | 0.05 | -0.01 | 1 | 1 |
| All | 0.27 | 0.23 | 0.23 | 0.24 | 0.04 | -0.01 | 98.82 | 98.82 |

[a] The abbreviation of different vegetation types are the same as it in Table 1; [b] Percentage means the percentage of sites that the reciprocal relationship was better than linear to regress the LUEc=GPP/PAR and PAR, i.e., with a higher $R^2$ and lower RMSE; [c] unit of RMSE is g C m$^{-2}$ hh$^{-1}$.

*4.2 Comparison of the parameterized maximum LUE*

Two parameters need to be optimized both in two models, i.e., $a$ and $\varepsilon^*_{max}$ in RTL-LUE and $\varepsilon_{msu}$ and $\varepsilon_{msh}$ in TL-LUE. The mean values, standard deviations (SD), and CV of them in different vegetation types are compared in Table 6. Overall, the values of $\varepsilon^*_{max}$ in the RTL-LUE model are lower than $\varepsilon_{msh}$ and higher than $\varepsilon_{msu}$ in the TL-LUE model among all the 12 vegetation types. In the TL-LUE model, the value of $\varepsilon_{msh}$ is much higher than $\varepsilon_{msu}$ because shaded leaves only have the chance to absorb the diffuse radiation, with $PAR_{sh}$ always much lower than $PAR_{su}$ and usually lies between the light compensation point and the light saturation point. According to the light response curve, the photosynthetic rate will decrease with increasing APAR, so the LUE of shaded leaves should be higher than that of sunlit leaves. Furthermore, the mean values of $\varepsilon_{msh}$ and $\varepsilon_{msu}$ are also compared with $\varepsilon_{max}^*$, which shows a very similar variation among different vegetation types with an $R^2$=0.73. The mean values of $\varepsilon_{msh}$ and $\varepsilon_{msu}$ are generally lower than $\varepsilon_{max}^*$, which may be due to the reason that shaded leaves are always more than sunlit leaves in a canopy. Since the value of $\varepsilon_{msh}$ is generally higher than that of $\varepsilon_{msu}$, the arithmetic mean of $\varepsilon_{msh}$ and $\varepsilon_{msu}$ with the same weight should be lower than the actual maximum LUE, in which the shaded leaves have a higher weight than sunlit leaves. So the variation of $\varepsilon_{max}^*$ among different vegetation types are also more similar to that of $\varepsilon_{msh}$ with $R^2$=0.79, but $R^2$ is only 0.34 for $\varepsilon_{msu}$.

**Table 6.** Mean value, Standard Deviations (SD), and CV of the optimized parameter of $a$ and different maximum LUE, i.e. $\varepsilon_{max}^*$, $\varepsilon_{msu}$, and $\varepsilon_{msh}$, for different vegetation types.

| Vegetation type [a] | Mean (g C MJ$^{-1}$) [b] | | | | SD (g C MJ$^{-1}$) [b] | | | | CV [c] (%) | | | |
|---|---|---|---|---|---|---|---|---|---|---|---|---|
| | $a$ | $\varepsilon_{max}^*$ | $\varepsilon_{msu}$ | $\varepsilon_{msh}$ | $a$ | $\varepsilon_{max}^*$ | $\varepsilon_{msu}$ | $\varepsilon_{msh}$ | $a$ | $\varepsilon_{max}^*$ | $\varepsilon_{msu}$ | $\varepsilon_{msh}$ |
| DBF | 1.54 | 2.85 | 0.78 | 3.13 | 1.08 | 0.91 | 0.31 | 0.87 | 69.90 | 32.08 | 39.59 | 27.82 |
| DNF | 0.69 | 1.68 | 0.70 | 1.89 | / | / | / | / | / | / | / | / |
| EBF | 1.02 | 2.16 | 0.72 | 2.40 | 0.76 | 0.83 | 0.12 | 0.60 | 74.37 | 38.46 | 16.65 | 25.06 |
| ENF | 0.79 | 2.14 | 0.88 | 2.57 | 0.51 | 0.58 | 0.32 | 0.74 | 64.79 | 27.07 | 36.99 | 28.91 |
| MF | 0.81 | 2.54 | 1.23 | 2.86 | 0.40 | 0.76 | 1.17 | 0.61 | 49.35 | 30.05 | 95.37 | 21.52 |
| GRA | 0.63 | 2.39 | 1.07 | 3.02 | 0.53 | 1.39 | 0.54 | 1.89 | 84.12 | 58.01 | 50.59 | 62.47 |
| CRO | 1.43 | 3.32 | 1.25 | 3.22 | 1.77 | 2.04 | 0.57 | 1.59 | 123.58 | 61.33 | 46.05 | 49.45 |
| CSH | 0.59 | 1.68 | 0.70 | 2.10 | / | / | / | / | / | / | / | / |
| OSH | 0.57 | 1.58 | 0.68 | 2.35 | 0.61 | 1.01 | 0.40 | 1.83 | 106.43 | 63.97 | 58.60 | 77.80 |
| WET | 1.36 | 2.82 | 1.14 | 3.26 | 2.18 | 1.85 | 0.66 | 2.36 | 159.96 | 67.08 | 57.63 | 72.49 |
| SAV | 0.20 | 1.60 | 1.41 | 1.91 | 0.00 | 1.10 | 0.98 | 1.16 | 0.00 | 66.85 | 69.57 | 60.93 |
| WSA | 0.46 | 1.96 | 0.97 | 3.01 | 0.31 | 0.70 | 0.27 | 1.57 | 68.14 | 31.95 | 28.27 | 52.01 |

[a] The abbreviations of different vegetation types are the same as it in Table 1; [b] the unit is only for $\varepsilon_{max}^*$, $\varepsilon_{msu}$, and

$\varepsilon_{msh}$, and the $PPFD_a$ is a parameter without unit; [c] CV=100%×SD/Mean.

The ranks of $\varepsilon_{msh}$ and $\varepsilon_{msu}$ among different vegetation types are consistent with the results of Zhou et al. (2016), in which the highest values are found in crops and the lowest values are in savannas. In this study, because three more vegetation types are included, $\varepsilon_{msu}$ of crops is the highest and $\varepsilon_{msu}$ of crops is only lower than that of wetland. The values in savannas are also the second-lowest and only slightly higher than those of DNF. For $\varepsilon_{max}^*$, the highest value is also found in crops, and the lowest value is in closed shrubs. There are some differences between the mean values of $\varepsilon_{msu}$ and $\varepsilon_{msh}$ in this study and Zhou et al. (2016) for overlapping nine vegetation types. The main reason is that there are many more sites employed to optimize the parameters in this study, in order to verify the reliability of the model across as many sites as possible. Zhou et al. (2016) parameterized the TL-LUE model by carefully excluding the flux sites with heterogeneous vegetation types. The accuracy of GPP simulations based on the suggested parameters for the nine vegetation types in Zhou et al. (2016) is also evaluated, and the results are very similar to those in this study, with a difference generally lower than 0.01 in $R^2$, as shown in Table S3. Because there are three more vegetation types used in this study, we do not apply the parameters suggested in Zhou et al. (2016) but optimize them ourselves for a fair comparison with the improved RTL-LUE model.

The parameter of $a$ also shows significant differences among different vegetation types. The highest $a$ values are found in the EBF and crops, and the lowest values are in SAV and WSA. Due to the reciprocal scalar of PPFD used in this study, the larger the value of $a$ for a vegetation type, the more sensitive to PAR the vegetation type is (Chen et al., 1999).

## 4.3 Sensitivity of GPP Simulations to PAR

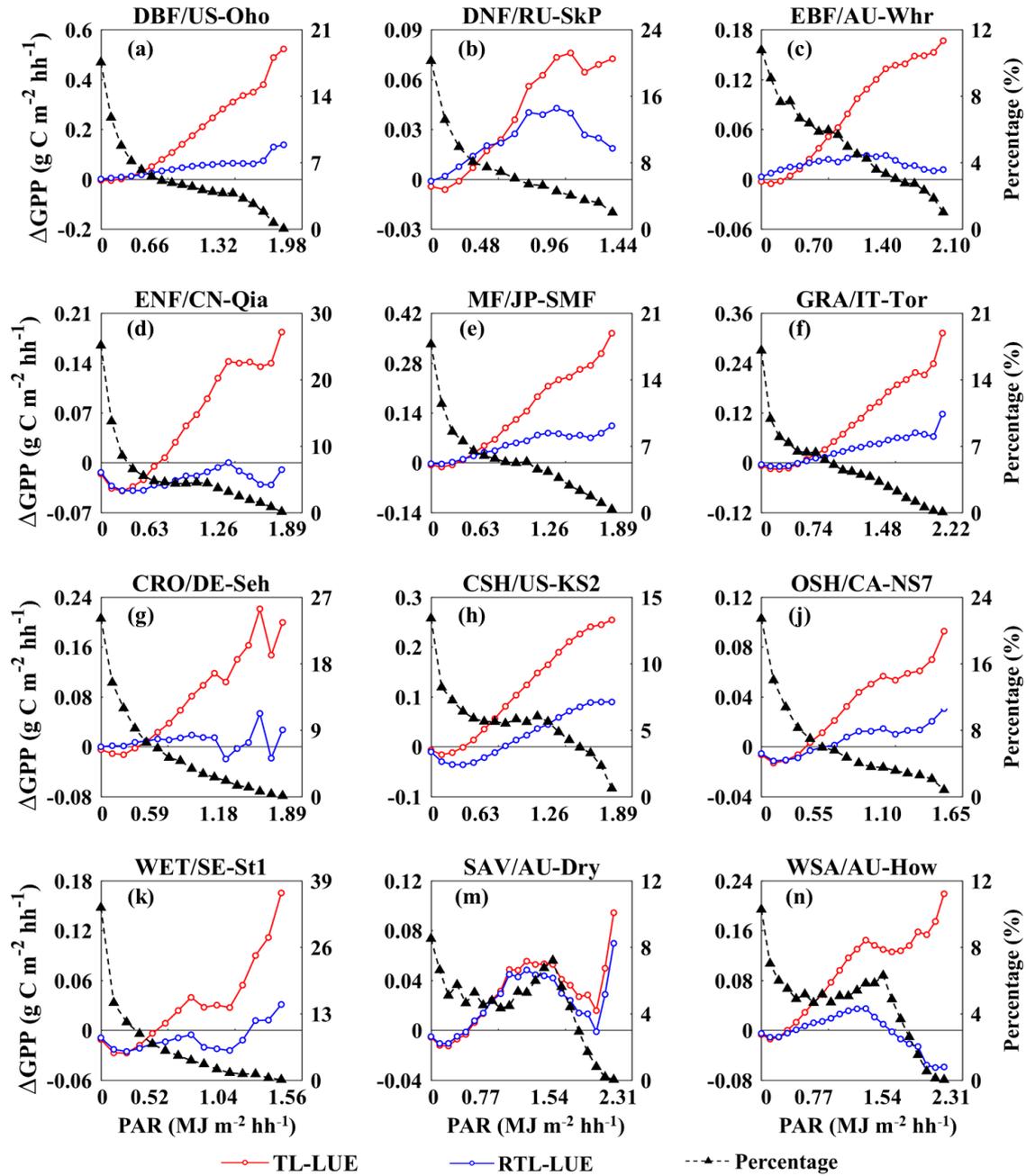

**Figure 8.** The sensitivity of GPP simulations to PAR for 12 specific sites with different vegetation types ( $\Delta GPP = GPP_{simulation} - GPP_{EC}$ , Percentage means the percentage of point numbers within the ranges of PAR). The title of each subplot is named as "vegetation types/site name".

The comparison of GPP simulations with EC measurements indicates that the TL-LUE model tends to overestimate GPP under the conditions of high incoming radiation. In order to further investigate the degree of overestimation, the difference of GPP simulations and EC GPP ($\Delta \text{GPP} = \text{GPP}_{simulation} - \text{GPP}_{EC}$) are binned according to the magnitude of incoming PAR. As shown in Fig. 8, both models tends to underestimate GPP in the conditions of very low PAR, and gradually overestimate GPP at high PAR values. GPP simulations from the two models can agree well with the EC GPP with little differences in the low incoming PAR range. While, with the increase in PAR, the overestimation of the TL-LUE model generally becomes more and more severe and reaches a very high level. In contrast, although overestimation of the RTL-LUE model can also be observed in some vegetation types, the magnitude of overestimation is much smaller and becomes stable even the PAR is very high. The RTL-LUE model dramatically alleviates the sensitivity of simulated GPP to PAR, by suppressing the high GPP to be more close to the EC measurements in the conditions of high incoming PAR. The difference between the two models at SAV sites is still very small, which is consistent with the results above due to the small value of $a$.

The main reason for the overestimation of GPP by the TL-LUE model in conditions of high incoming PAR is the use of constant LUE values without scaling for radiation. The percentage of point numbers within different PAR ranges is also calculated and displayed as the black triangular line in Fig. 8. Obviously, much more points are concentrated in the area with low incoming PAR, where the GPP simulated by the TL-LUE model agrees well with the EC GPP. As a result, the parameter optimization tends to obtain a relatively high value of $\varepsilon_{msu}$ and $\varepsilon_{msh}$ that suit for the majority conditions of low incoming PAR, in order to ensure the overall accuracy. Besides, because only the scalars of VPD and temperature are considered in the TL-LUE model, the LUEs of sunlit and shaded leaves will not directly change with PAR. So the constantly high $\varepsilon_{msu}$ and $\varepsilon_{msh}$ will certainly induce the unreasonable high LUE, resulting in the overestimation of GPP in the conditions of high incoming PAR. This problem can be significantly alleviated in the RTL-LUE model, because the scalar of incoming PAR is used to suppress the LUE at higher incoming PAR.

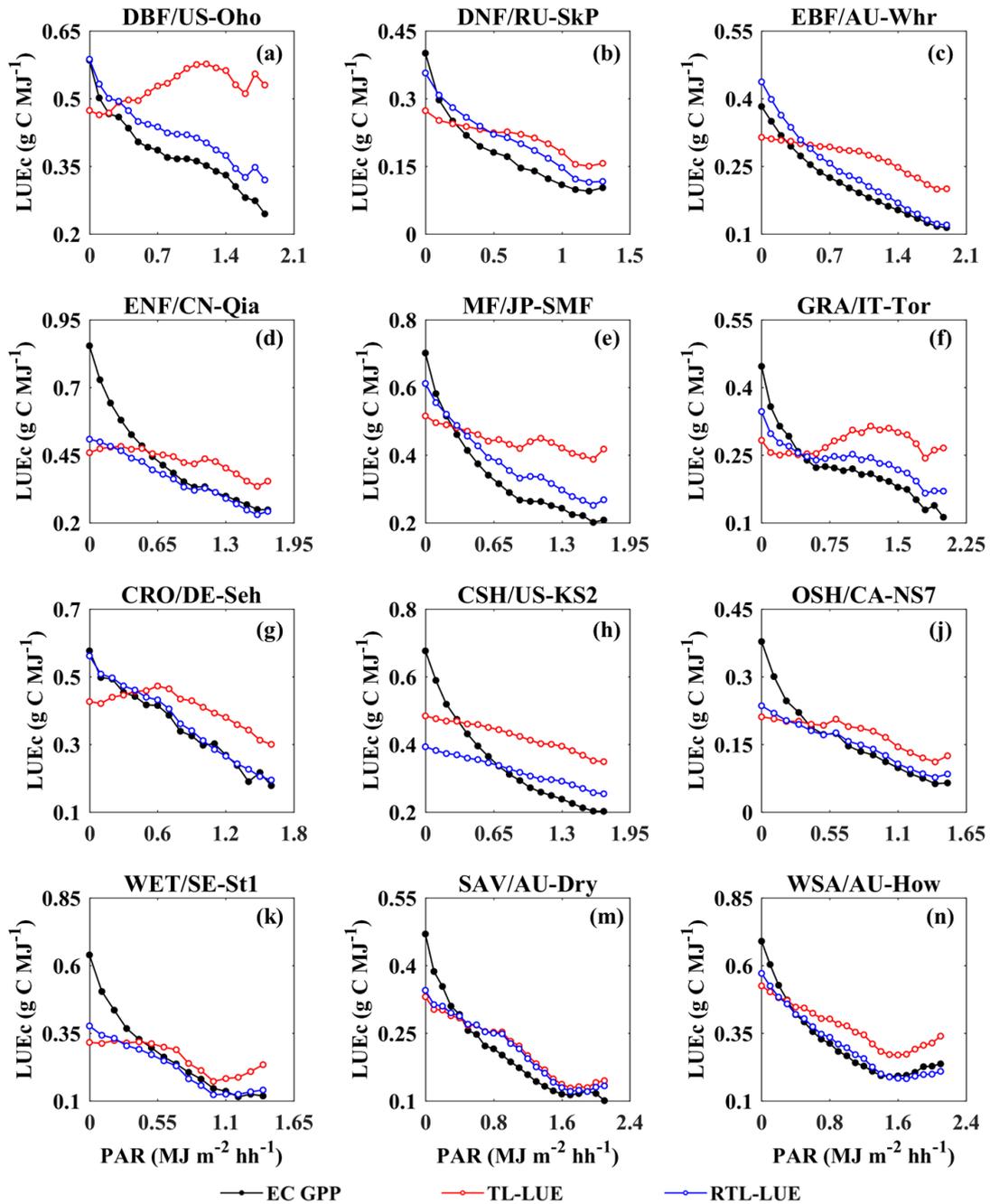

**Figure 9.** Comparison of the dependence of canopy light use efficiency (LUEc = GPP/PAR) on PAR averaged over 0.01 bins of PAR at 12 specific sites with different vegetation types. The title of each subplot is named as "vegetation types/site name".

The canopy LUEs (LUEc=GPP/PAR) calculated by GPP simulations and EC observations are also averaged over 0.01 bins of incoming PAR (MJ m$^{-2}$ hh$^{-1}$), as shown in Fig. 9. At all sites, the EC observed LUEc significantly decreases with PAR, and the RTL-LUE model can generally reproduce these trends. However, the variation of LUEc simulated by the TL-LUE model shows more flat trends. Obviously higher LUEc values in the TL-LUE model can be observed at high incoming PAR, indicating that the maximum LUE could not be reduced to the actual LUE only by the scalars of VPD and temperature in the TL-LUE model, and thus leading to the overestimation of GPP at high incoming PAR. It is reasonable to consider the constraint of radiation in the RTL-LUE model, and the overestimation could be significantly alleviated by *f(PPFD)*.

## *4.4 Improvements and limitations*

In this study, an RTL-LUE model is developed to improve GPP simulation by considering the scalar of radiation on vegetation LUE, based on the concept that LUE should decrease with increasing incoming PAR (Chen et al., 1999; Koyama and Kikuzawa, 2010; Sinclair and Horie, 1989). The model is modified based on a TL-LUE model by giving the same maximum LUE to the sunlit and shaded leaves, rather than prescribing two different maximum LUE values for these two leaf groups. Generally, the maximum LUE, which is controlled by physiological traits of leaves in a plant canopy, should be almost the same for shaded or sunlit leaves (Leverenz, 1987). Although limited differences may exist in the distributions of nutrients and leaf age between sunlit and shaded leaves, their impacts on the maximum LUE are expected to be small (Hikosaka, 1996). The major differences between sunlit and shaded leaves are the environmental conditions, i.e., radiation, temperature, and humidity, which will only impact the actual LUE rather than maximum LUE. Among these environmental variables, radiation is the primary factor because the radiation absorbed by shaded leaves is undoubtedly much less than that by sunlit leaves (McCallum et al., 2013; Perkins et al., 2006). The results and analysis above have shown the effectiveness of the scalar of radiation in constraining the maximum LUE and representing the differences in LUE between sunlit and shaded leaves. After modifying the model, no additional data inputs are needed, and the number of parameters needed to optimize is also the same, which are $\varepsilon_{msu}$ and $\varepsilon_{msh}$ for the TL-LUE model and $a$ and $\varepsilon_{max}^*$ for the RTL-LUE model. If we take $a$ as the fixed-parameter for environmental scalars, similar to $VPD_{max}$, $VPD_{min}$, $T_{max}$, or $T_{min}$, only $\varepsilon_{max}^*$ is needed to optimize in the RTL-LUE model.

However, there still exist some limitations in this study. First of all, LAI is the core inputs for both the TL-LUE and RTL-LUE models, and the use of remote sensing LAI may induce uncertainties in parameter calibration and the GPP validation (Zhou et al., 2016). Although the GLASS LAI product used in this study has been shown to have satisfactory performance in simulating GPP (Xiao et al., 2017; Xie et al., 2019), there are still errors in the data. Cloud and atmospheric conditions will cause some abnormal variations in the LAI time series, so it cannot always well interpret the vegetation structure variation even after the time filtering process. It may be one of the main reasons for the unsatisfactory results in crops, such as the LAI changes in the harvest period are not well interpreted and still retain a high value. As a result, both models failed to capture the seasonal variation of crop GPP during these periods. Besides, the mismatch of the spatial resolution of LAI products and the flux footprints may also lead to some uncertainties (Chen, 1999; He et al., 2013). LAI of a pixel from the GLASS product represents the mean value of the 1 km area, rather than the mean LAI of the unique vegetation types or regions measured by a flux tower. Further studies can be conducted using LAI products with high spatial resolutions and high accuracy in conjunction with flux footprints. In addition, the EC sites in the FLUX2015 dataset with more than one-year of valid half-hour data are all selected in this study. The neglect of the heterogeneous land cover of the flux sites may also induce uncertainties both in the parameterization and validation (Zhou et al., 2016). The low criteria are employed to select data because we want to verify the accuracy of the model as broad as possible. Further research can be conducted to explore whether there are differences in the conclusions if the flux sites with very heterogeneous vegetation types are excluded. Furthermore, we assume that the discrepancies in LUE between sunlit and shaded leaves are mainly decided by the different light intensity they received, and the impacts of temperature, humidity, and nutrients are ignored. Actually, the sunlit leaves are supposed to have higher temperatures because they are directly exposed to the full sunlight, which will also impact the water vapor deficit at the leaf surface (Gates, 1964; Leuning et al., 1995; Miller, 1971). As a result, it is interesting to investigate if the temperature and humidity scalars for sunlit and shaded leaves can be calculated differently. These scalars used in the RTL-LUE model are the same as the MOD17 algorithm. The exclusion of soil water content, $CO_2$ fertilization, and nutrient conditions in optimizing LUE may have also limited the accuracy of GPP estimation to some extent (Chen et al., 2019a; Wang et al., 2020). Further efforts are needed to synthesizing the impacts of these factors on LUE.

## 5. Conclusion

In this study, a radiation-constrained two-leaf light use efficiency (RTL-LUE) model is developed to improve the simulation of GPP on the basis of a TL-LUE model. This new model assigns the same maximum LUE to both sunlit and shaded leaves, and the differences in LUE between them are represented by a scalar of radiation. The core parameters of TL-LUE and RTL-LUE models were optimized and the accuracies of GPP simulations by these two models were assessed based on the measurements from globally distributed 169 EC sites. GPP simulations from the two models generally agree well with EC GPP, but consistent improvements can be observed in results from RTL-LUE ($R^2$=0.74, RMSE=358.81 g C m$^{-2}$ year$^{-1}$) over TL-LUE ($R^2$=0.72, RMSE= 524.09 g C m$^{-2}$ year$^{-1}$). The improvements are the largest at forest sites. GPP simulations from RTL-LUE show higher $R^2$ and lower RMSE and bias than those from TL-LUE either in all or individual vegetation types, for half-hour, 8-day, and yearly time scales. Besides, the diurnal and seasonal variations of GPP simulated by RTL-LUE also fit better with the EC observations than those by TL-LUE, because the overestimation at high incoming PAR (at noon and in peak growing season) by TL-LUE is significantly alleviated by the radiation scalar used in RTL-LUE. The optimized values of $\varepsilon_{max}^*$ in the RTL-LUE model are generally lower than $\varepsilon_{msh}$ and higher than $\varepsilon_{msu}$ optimized in the TL-LUE model, and show similar variation with $\varepsilon_{msh}$ among 12 vegetation types. The scatter plots of canopy LUE and PAR based on EC observations show that it is reasonable to use the reciprocal scalar of radiation to constrain LUE. Further statistics indicate that the RTL-LUE model showed much lower sensitivity of error to PAR after considering the scalar of radiation. The overestimation of GPP in the TL-LUE model can be mainly explained by the optimized radiation-independent constant $\varepsilon_{msh}$ and $\varepsilon_{msu}$ values, which are only suitable for low-medium incoming PAR conditions that occurred most frequently and had the largest influence on the optimization of $\varepsilon_{msh}$ and $\varepsilon_{msu}$.

This study highlights the importance of the radiation scalar in simulating GPP in LUE models even after the differentiation of LUE between sunlit and shaded leaf groups, because LUE would still vary within the same group and the difference in LUE between the two groups would also vary with radiation. We also demonstrate the excellent performance of the RTL-LUE model under most circumstances without differentiating the maximum LUE between sunlit and shaded leaves in order to minimize the number of parameters to be optimized. This seems to be a workable strategy as the small difference in leaf physiology between sunlit and shaded leaves would be superseded by the difference in light intensity on these two leaf groups. Further efforts in improving the RTL-LUE models may be directed towards the scalars of temperature and VPD and the possible differentiation of these scalars between sunlit and shaded leaves.

## Acknowledgments

This research was supported by the National Key Research and Development Program of China (2019YFB2102903), the National Natural Science Foundation of China (42001371), and the China Scholarship Council. We are grateful to all data providers, including the FLUXNET community and GLASS LAI team. Special thanks are given to all the PIs who contributed to the FLUXNET2015 dataset and provided valuable data for the research community. All the flux and meteorological data are available from the website (https://fluxnet.org/data/fluxnet2015-dataset/), which are acquired and shared by global collaborated regional networks supporting the FLUXNET.